\documentclass[10pt,twocolumn,letterpaper]{article}

\newcommand\blfootnote[1]{%
  \begingroup
  \renewcommand\thefootnote{}\footnote{#1}%
  \addtocounter{footnote}{-1}%
  \endgroup
}

\usepackage{cvpr}              

\usepackage{booktabs}   
\usepackage{graphicx}

\usepackage{amsmath}
\usepackage{multirow} 
\usepackage{caption}
\usepackage{subcaption} 

\usepackage[accsupp]{axessibility}
\definecolor{cvprblue}{rgb}{0.21,0.49,0.74}
\usepackage[pagebackref,breaklinks,colorlinks,allcolors=cvprblue]{hyperref}

\newcommand{\lfl}[1]{{\color[rgb]{0., 0., 0.}#1}}

\newcommand{\kjh}[1]{{\color[rgb]{0, 0, 0}#1}}

\makeatletter
\renewcommand{\paragraph}{%
  \@startsection{paragraph}{4}%
  {\z@}{0.2ex \@plus 0.3ex \@minus .2ex}{-1em}%
  {\normalfont\normalsize\bfseries}%
}
\makeatother
\def\paperID{42281} 
\def\confName{CVPR}
\def\confYear{2026}

\title{SketchFaceGS: Real-Time Sketch-Driven Face Editing and Generation with Gaussian Splatting}

\author{
    Bo Li\textsuperscript{1,2} \quad 
    Jiahao Kang\textsuperscript{2} \quad 
    Yubo Ma\textsuperscript{2} \quad 
    Feng-Lin Liu\textsuperscript{3,4} \quad 
    Bin Liu\textsuperscript{2} \quad 
    Fang-Lue Zhang\textsuperscript{5} \quad 
    Lin Gao\textsuperscript{3,4}\thanks{Corresponding author.} 
    \vspace{1.5mm} \\
    \textsuperscript{1}Shandong Technology and Business University \\
    \textsuperscript{2}Nanchang Hangkong University \\
    \textsuperscript{3}Institute of Computing Technology, Chinese Academy of Sciences \\
    \textsuperscript{4}University of Chinese Academy of Sciences \\
    \textsuperscript{5} University of New South Wales 
    \vspace{1.5mm} \\
    {\tt\small bolimath@gmail.com, 2307070100013@stu.nchu.edu.cn, 2304081200005@stu.nchu.edu.cn,} \\
    {\tt\small liufenglin21s@ict.ac.cn, nyliubin@nchu.edu.cn, z.fanglue@gmail.com, gaolin@ict.ac.cn}
}
\begin{document}

\twocolumn[{
\maketitle
\begin{center}
    \includegraphics[width=\textwidth]{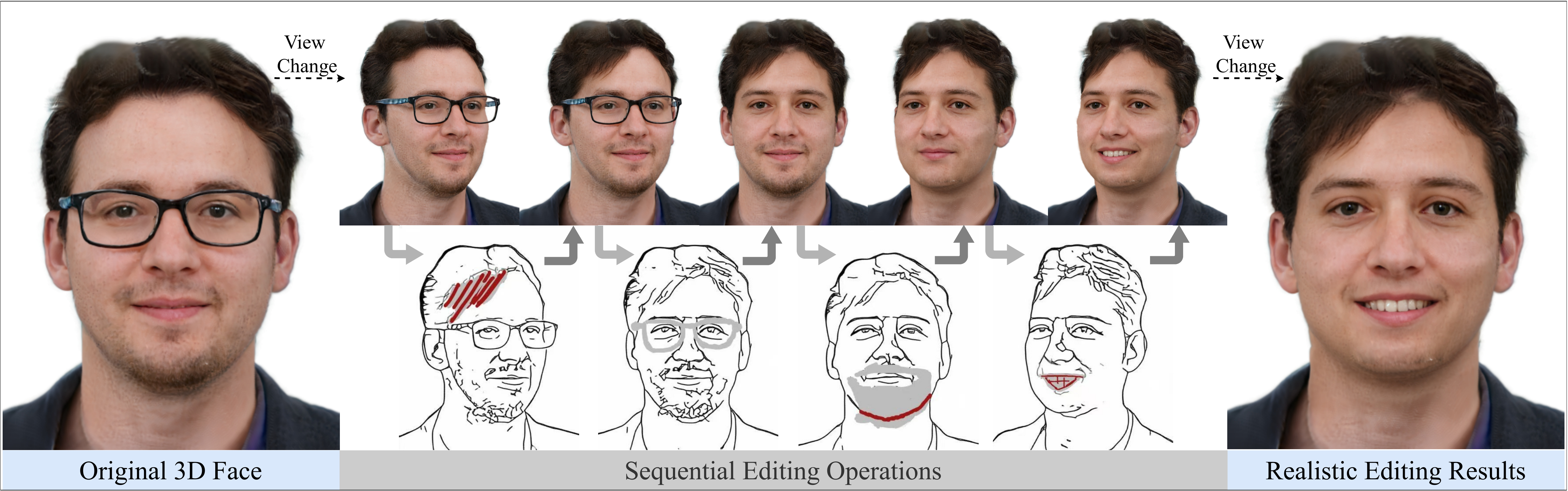}
    \captionof{figure}{
    \lfl{SketchFaceGS presents a real-time, sketch-driven editing framework tailored for 3D Gaussian Splatting (3DGS) facial heads. Starting with an original 3D face (a), sequential editing operations (b) are applied to targeted local regions. Our approach delivers photo-realistic editing results (c) while preserving the 3D consistency.}
    }
\end{center}
}]

\blfootnote{* Corresponding author.}

\maketitle
\begin{abstract}

    3D Gaussian representations have emerged as a powerful paradigm for digital head modeling, achieving  photorealistic quality with real-time rendering. However, intuitive and interactive creation or editing of 3D Gaussian head models remains challenging. Although 2D sketches provide an ideal interaction modality for fast, intuitive conceptual design, they are sparse, depth-ambiguous, and lack high-frequency appearance cues, making it difficult to infer dense, geometrically consistent 3D Gaussian structures from strokes—especially under real-time constraints. To address these challenges, we propose SketchFaceGS, the first sketch-driven framework for real-time generation and editing of photorealistic 3D Gaussian head models from 2D sketches.
    Our method uses a feed-forward, coarse-to-fine architecture. A Transformer-based UV feature-prediction module first reconstructs a coarse but geometrically consistent UV feature map from the input sketch, and then a 3D UV feature enhancement module refines it with high-frequency, photorealistic detail to produce a high-fidelity 3D head. For editing, we introduce a UV Mask Fusion technique combined with a layer-by-layer feature-fusion strategy, enabling precise, real-time, free-viewpoint modifications. Extensive experiments show that SketchFaceGS outperforms existing methods in both generation fidelity and editing flexibility, producing high-quality, editable 3D heads from sketches in a single forward pass.

\end{abstract}    

\section{Introduction}
\label{sec:intro}

Accurate 3D facial modeling remains a challenging problem in computer graphics and computer vision, underpinning applications from cinematic characters to immersive avatars. Prior mesh‑based pipelines~\cite{Portenier2018,Maya2019} typically rely on labor‑intensive workflows to produce photorealistic faces that capture both geometry and appearance, making them impractical for non‑experts and costly to achieve photoreal quality. Recent NeRF-based approaches~\cite{Bergman,Gu2021,Niemeyer_2021,Schwarz2020} leverage 3D GANs for controllable face synthesis; however, their practical adoption is often limited by costly rendering and 3D inconsistencies introduced by super-resolution modules. Although 3D Gaussian splatting~\cite{gaussian_splatting} addresses real‑time rendering, existing GAN‑based Gaussian head generation methods~\cite{Hierarchical_gan,gghead} and text‑driven face‑editing approaches~\cite{gaussianeditor,Vachha2024,Wu2024} lack fine‑grained, convenient control for detailed facial edits.

We propose an intuitive, sketch-driven approach for 3D face editing that leverages 3D Gaussian Splatting. To overcome \kjh{the reliance on time-consuming per-instance optimization} and error accumulation in existing 3D sketch-based methods (e.g., SketchFaceNeRF~\cite{gao2023sketchfacenerf}), we introduce a unified, \kjh{optimization-free}, feed-forward framework that directly generates and interactively edits photorealistic 3D Gaussian head models from single-view hand-drawn sketches. Our method bridges the gap between sparse 2D sketches and dense 3D Gaussian representations via a coarse-to-fine pipeline. In the coarse stage, two parallel Transformer branches extract geometric features from the sketch and appearance features from a reference image; a fusion network then reconciles geometry–appearance conflicts across identities to produce a geometry-consistent, color-coherent coarse UV map. In the fine stage, we use a pretrained 3D-GAN as a high-frequency texture prior: a U-Net maps the coarse UV into a global latent and multi-scale spatial modulation parameters, which are applied layer-wise to condition synthesis and elevate the coarse representation into a renderable, photorealistic 3D face. \kjh{Essentially, our proposed feed-forward architecture effectively bridges abstract sketches with powerful 3D generative priors, thereby achieving photorealistic, real-time, and geometrically consistent conditional generation.}

For interactive editing, we introduce UV Mask Fusion. User edits (draw/erase) are converted to 2D pixel masks and back-projected into 3D to identify the 3D Gaussians that contribute to the edited pixels. Mapping those Gaussians into the canonical UV map produces precise UV masks that separate edited and unedited regions. Instead of compositing in 3D Gaussian Splatting space, we resample each UV mask to match the resolution of the corresponding generator layer and perform layer-wise fusion in the generator’s multi-scale feature maps. This suppresses seam artifacts and yields natural, seamless merges. By leveraging the model’s end-to-end generative capability, our approach enables continuous, free‑viewpoint, real‑time editing, \kjh{while inherently preserving the structural integrity and identity of the original face in non-edited regions.}

Our main contributions are:
\begin{itemize}
  \item We introduce the first unified, end-to-end, \kjh{and optimization-free} framework to generate and interactively edit photorealistic 3D Gaussian head models from a single sketch, providing an intuitive, artist-friendly workflow.
  \item 
  \lfl{We propose a coarse-to-fine pipeline \kjh{that effectively bridges abstract sketches and 3D priors,} where parallel Transformer branches extract geometry and appearance from the sketch and reference image respectively to produce a coarse UV map. A 3D-GAN–based modulation module then enhances it with high-frequency, photorealistic details.}
  \item We present UV Mask Fusion, a real-time editing mechanism that performs layer-wise feature fusion in the generator’s multi-scale maps, \kjh{inherently preserving the identity of unedited regions and} avoiding spatial composition artifacts and enabling natural, stable, continuous free-viewpoint editing.
\end{itemize}

\section{Related Work}
\label{sec:related work}

\noindent\textbf{3D-aware Portrait Synthesis}
\label{3D Face Generation and Editing}
3D-GANs~\cite{gan,eg3d} are more popular in 3D-aware portrait synthesis compared with Diffusion Models~\cite{ho2020denoising,zhang2025magictalk} because of the training dependence on 2D images instead of scarce 3D data.
Early 3D-GAN 
methods that render pixels directly from 
3D representations, including meshes~\cite{Chen_Zhang_2019,Kanazawa_2018}, voxels~\cite{Gadelha_2016,Henzler_2019}, and 
Neural Radiance Fields~\cite{2021pi,Schwarz2020,sun2024implicit}\lfl{, with later works introducing additional convolution modules~\cite{Nguyen-Phuoc_2019,Niemeyer_2021,Xue} to improve fidelity and resolution}. 
Subsequent approaches~\cite{eg3d,Gu2021,OrEl_2022} adopted StyleGAN~\cite{Karras_2019} as the generator backbone, \kjh{facilitating 3D face manipulation~\cite{gao2025learning} and disentangled lighting representation~\cite{nerffacelighting}.}
Among them, EG3D~\cite{eg3d} proposed a tri-plane-based hybrid neural rendering with a super-resolution module for upsampling. However, its volume rendering remained inefficient, and the super-resolution stage introduced cross-view inconsistencies. 
To address these issues, recent work leverages 3DGS~\cite{gaussian_splatting,wu2024recent}, a new explicit representation \kjh{enabling native high-resolution, real-time head synthesis. Moreover, recent advances have extended 3D Gaussian representations to dynamic head avatar animation~\cite{3dgsblendshapes,rgbavatar,realtimehighfidelity} and to face reconstruction for virtual interaction~\cite{hmdgaze}. GSGAN~\cite{Hierarchical_gan} introduces a hierarchical multiscale Gaussian generator} that regularizes the Gaussians across levels, enabling coarse-to-fine 3D modeling.
GGHead~\cite{gghead} binds 3D Gaussians to a template mesh and predicts their attributes as UV maps within a StyleGAN framework, \lfl{achieving real-time and arbitrary resolution rendering.}
Our work 
builds on these latest advances in 3D Gaussian head generation.

\noindent\textbf{Sketch-based Face Modeling} 
\label{Neural Sketch-based Face Generation}
Sketching has been widely used for both facial image generation~\cite{chen2020deep,LinesToFacePhoto,DeepFacePencil,ControllableFace} and interactive editing/reshaping~\cite{chen2021deepfaceediting,Jo_Park_2019,yang2020deep,zeng2022sketchedit,chen2024deepfacereshaping}.
Some 3D face modeling methods predict the coefficients of parametric models and refine surface details with displacement maps~\cite{DeepSketch2Face,Han_2018,yang2021learning}, while others use sketches to guide the deformation of a template mesh to produce diverse 3D faces~\cite{du2020sanihead,Han_2018,luo2021simpmodeling}.
Although effective for mesh geometry, these approaches struggle to produce photorealistic images due to the difficulty of estimating texture, materials, and lighting.
To address this issue, recent research proposes NeRF-based generation methods.
For instance, S3D~\cite{song2025s3d} 
generates a semantic mask from a face sketch and synthesizes a 3D face using a semantic-map-based generative network like pix2pix3d~\cite{kangle2023pix2pix3d}, but this sketch-to-mask mapping loses fine-grained texture details.
SketchFaceNeRF~\cite{gao2023sketchfacenerf} predicts a Facial NeRF from a sketch using a tri-plane network and enables 3D face editing via mask fusion. However, its coarse tri-plane prediction reduces generation quality, and editing requires time-consuming per-instance optimization. Additionally, sequential edits sometimes cause error accumulation, preventing their use in real-time interactive facial creation.
In contrast, our model is a feed-forward framework that enables real-time, optimization-free editing, where precise UV Mask Fusion effectively prevents error accumulation.

\noindent\textbf{Feed-forward 3D Reconstruction Model} 
\label{Feed-forward Reconstruction Model}
\lfl{Neural representations, such as NeRF~\cite{Mildenhall2020} and 3DGS~\cite{gaussian_splatting}, have revolutionized 3D reconstruction with their high-quality novel view synthesis. To eliminate costly per-scene optimization, Large Reconstruction Models (LRMs)~\cite{hong2023lrm}} series 
introduced Transformer-based feed-forward architectures, enabling fast and generalizable reconstruction of detailed 3D models from sparse inputs. 
This powerful paradigm has rapidly extended from general object reconstruction to the domain of human body reconstruction. For instance, Human-LRM~\cite{weng2024template} utilizes a Transformer to decode tri-plane-NeRF representations, while Human-Splat~\cite{pan2024humansplat} employs a latent reconstruction Transformer to directly predict 3DGS human avatars from multi-view images generated by a diffusion model.
In the more fine-grained task of head modeling, 
GAGAvatar~\cite{chu2024gagavatar} proposed a one-shot, 3D Gaussian-based method for head avatar generation, \lfl{though it still requires 2D neural post-processing for ideal animation.}
In contrast, LAM~\cite{lam} builds Gaussian features on FLAME vertices and densifies them via subdivision and interpolation to achieve fast and realistic 3D head reconstruction.
Inspired by LAM's efficient feed-forward architecture, our model uses a Transformer-based feature extraction network, aiming to \lfl{improve the head’s realism and its structural consistency with the input sketches.}

\section{Method}
\label{sec:method}

\begin{figure*}[t]
    \centering
    \includegraphics[width=1\textwidth]{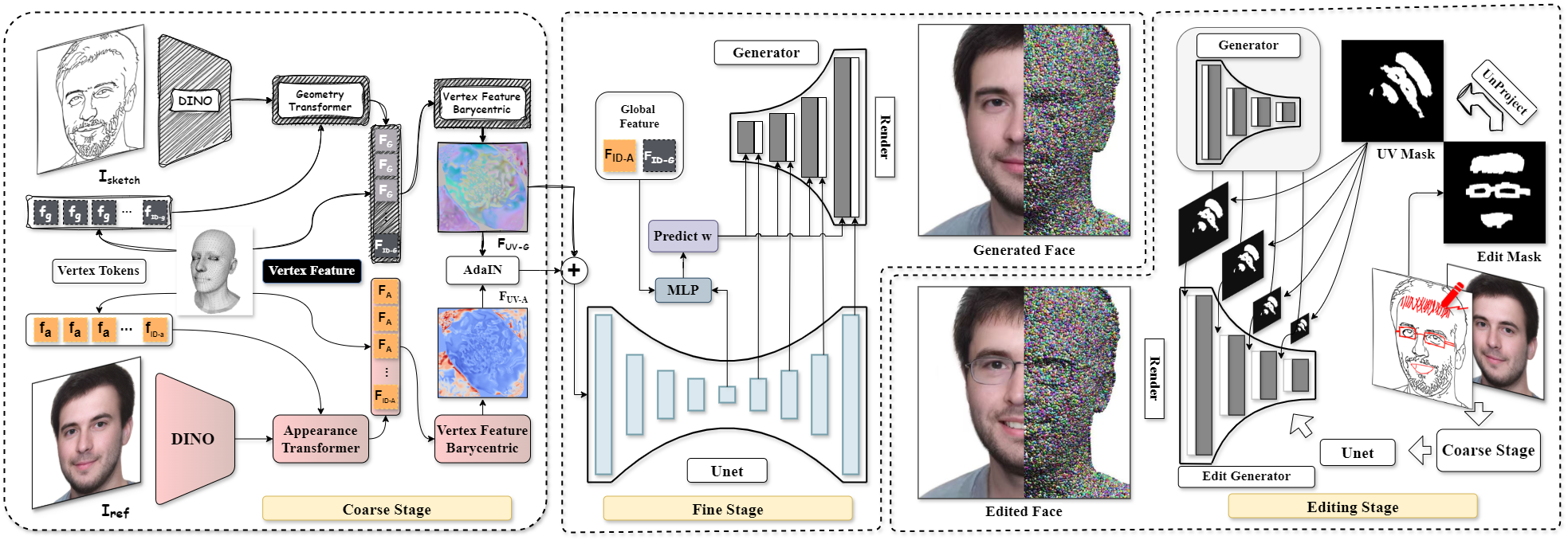}
    \caption{\textbf{Overview of the SketchFaceGS Framework.} Our method consists of two core components: (a) a sketch-based generation pipeline that translates a 2D sketch and a reference image into a photorealistic 3D Gaussian head through a coarse-to-fine process, and (b) a real-time editing pipeline that leverages a novel UV Mask Fusion method and a layer-wise feature fusion strategy to enable precise, continuous, and view-independent modifications. 
    }
    \label{fig:pipeline}
\end{figure*}

We propose a real-time 3D head generation and editing framework driven by sketches. As shown in Fig.~\ref{fig:pipeline}, given a single portrait image and a hand-drawn sketch as inputs, our method enables intuitive and efficient manipulation of 3D head models by simply editing the sketch.
Section 3.1 introduces some prerequisite knowledge related to our method. Section 3.2 provides a detailed description of the steps involved in our method during the generation phase, whereas the editing component is thoroughly explained in Section 3.3. Section 3.4 elaborates on the loss at each stage.

\subsection{Preliminaries}
\label{sec:preliminaries}

\paragraph{3D Gaussian Splatting.}
Our work is based on 3D Gaussian Splatting~\cite{gaussian_splatting}, a point-based 3D representation that assigns 5 core attributes to each point: position $\mu \in \mathbb{R}^3$, scale $s \in \mathbb{R}^3$, rotation quaternion $q \in \mathbb{R}^4$, opacity $\alpha \in \mathbb{R}$, and color $c \in \mathbb{R}^3$. A Gaussian ellipsoid is formulated as:
\begin{equation}
    G(x) = e^{-\frac{1}{2}(x-\mu)^\mathrm{T}\Sigma^{-1}(x-\mu)}, \quad \text{with } \Sigma = RSS^\mathrm{T}R^\mathrm{T}
\end{equation}
where $x$ is the world coordinates, and the scaling matrix $S$ and rotation matrix $R$ are derived from the scaling $s$ and quaternion $q$. The color $C$ of a pixel in image space is computed by blending $N$ overlapping sorted points:
\begin{equation}
    C = \sum_{i=1}^{N} c_i \alpha_i \prod_{j=1}^{i-1} (1 - \alpha_j)
\end{equation}
This representation enables high-fidelity, real-time rendering, making it suitable for interactive facial content creation.
\paragraph{Generative Prior for 3D Gaussian Heads.}
To leverage the power of generative models, we adopt GGHead~\cite{gghead} as our core generative prior. GGHead employs a StyleGAN2~\cite{stylegan2} backbone that maps a latent code $\mathbf{w} \in \mathcal{W}^{+}$ to a multi-channel UV map. This map directly parameterizes the attributes of a full 3D Gaussian head. In our framework, we repurpose GGHead's pre-trained StyleGAN backbone not for random generation, but as a powerful and controllable decoder that translates our predicted UV features into a high-fidelity 3D head model.
\subsection{Sketch-based Gaussian Head Generation}
\label{sec:generation}
Directly mapping a sparse sketch to a 3D head is highly ill-posed due to the inherent ambiguity between sparse line structures and dense 3D geometry. To progressively resolve this, we design a feed-forward, coarse-to-fine architecture: although the coarse stage predicts a UV feature map to establish a geometrically consistent foundation, it primarily captures low-frequency structural and chromatic information. The subsequent fine stage refines this representation via a GAN-based UV feature enhancement module that injects high-frequency, photorealistic details.

\subsubsection{Coarse Stage: Transformer-based UV Prediction}
\label{sec:coarse_stage}

Given a sketch $I_{\text{sketch}}$ and a reference $I_{\text{ref}}$, we predict a coarse but geometry-consistent UV feature map. To disentangle geometric and appearance information, we employ a parallel Transformer architecture with distinct branches for 
distinct inputs.


Inspired by the 3D-aware querying mechanism in LAM~\cite{lam}, we employ a Transformer backbone for sketch-to-geometry prediction. In this framework, a set of learnable queries, corresponding to the vertices of a canonical head template, interacts with features extracted from the input sketch via cross-attention to predict per-vertex geometric attributes. A similar, smaller Transformer is used to extract appearance features from $I_{ref}$. More specifically, we first extract deep features $F_{sketch}$ and $F_{ref}$ from the input sketch and color image using a pretrained DINOv2~\cite{oquab2023dinov2} encoder.
These image features serve as the keys and values in each Transformer's cross-attention mechanism, which operates on learnable vertex queries $f_{\text{g}}$, $f_{\text{a}}$ and global identity tokens $f_{{\text{ID-g}}}$, $f_{{\text{ID-a}}}$ to produce both per-vertex features $F_{\text{G}}$, $F_{\text{A}}$ and global identity vectors $F_{\text{ID-G}}$, $F_{\text{ID-A}}$:
\begin{align}
    F_{\text{G}}, F_{\text{ID-G}} &= \mathbf{T}_{\text{G}}((f_{\text{g}}, f_{\text{ID-g}}), F_{\text{sketch}}) \\
    F_{\text{A}}, F_{\text{ID-A}} &= \mathbf{T}_{\text{A}}((f_{\text{a}}, f_{\text{ID-a}}), F_{\text{ref}})
\end{align}
We apply barycentric interpolation to project these vertex features $F_{\text{G}}$, $F_{\text{A}}$ onto the UV map, obtaining dense $F_{\text{UV-G}}$ and $F_{\text{UV-A}}$. The $F_{\text{ID-G}}$ and $F_{\text{ID-A}}$ features are utilized in the fine stage to preserve the human identity.


Notably, independent feature extraction may clash if the structures of sketch and reference differ significantly.
To resolve this, we introduce a AdaIN-based alignment network $G_c$ to align these two representations.
Specifically, the network 
\lfl{normalizes the geometric feature map $F_{\text{UV-G}}$, then applies scaling and biasing using the corresponding scalar components from appearance feature map $F_{\text{UV-A}}$}, producing a new, coherent feature map:
\begin{equation}
F_{\text{UV-align}} = G_c(F_{\text{UV-G}}, F_{\text{UV-A}})
\end{equation}
Finally, the aligned appearance map $F_{\text{UV-align}}$ is concatenated with the geometric map $F_{\text{UV-G}}$ to form a coherent input for the subsequent fine stage.




\subsubsection{Fine Stage: 3D UV Feature Enhancement}
\label{sec:fine_stage}

The coarse UV map from the previous stage establishes the foundational geometry and base color but lacks photorealistic detail.
To inject high-frequency textures, we introduce a UV feature enhancement module. 
This module adapts the concept of generative modulation, 
originally popularized in 2D face restoration 
models such as GFP-GAN~\cite{gfp-gan}, and applies it to our 3D UV feature space.

Specifically, we design a U-Net that takes the coarse UV feature map as input and predicts a set of modulation parameters for our pre-trained GGHead generator. These parameters consist of a global feature vector $F_{latent}$ and a pyramid of multi-resolution spatial features $F_{spatial}$. 
 The feature vector $F_{latent}$ is then aggregated with the geometric and appearance identity vectors ($F_{\text{ID-G}}, F_{\text{ID-A}}$) extracted during the coarse stage. An MLP projects this combined representation into the StyleGAN's $\mathcal{W}^{+}$ latent space to produce the final identity-aware latent code $\mathcal{W}$:
\begin{equation}
    \mathcal{W} = \mathrm{MLP}\bigl(\mathrm{concat}(F_{\text{latent}}, F_{\text{ID-G}}, F_{\text{ID-A}})\bigr)
\end{equation}
The predicted features guide the StyleGAN generator via two complementary mechanisms—global identity injection and localized detail modulation—to synthesize a final, detail-rich UV map, $F_{output}$. This map directly encodes the complete attributes for the 3D Gaussian primitive set, which are then rendered to generate the final photorealistic image.

\subsection{Sketch-based Gaussian Head Editing}
\label{sec:editing}
Our sketch-driven geometry representation enables intuitive, interactive facial editing within the generative pipeline. Leveraging the learned features from the edited sketch, we support real-time, continuous edits to the 3D Gaussian head, including shape adjustments (e.g., facial contours) and local feature edits (e.g., eyes and hair), producing consistent, photorealistic updates to the geometry and appearance of the generated head model.
The process includes accurate localization of edits from 2D screen space to canonical UV space and coherent fusion of new and original content within the generator's feature space. The pipeline comprises two key components: UV-mask localization and layer-wise feature fusion.

\noindent\textbf{UV Mask Synthesis.}
\lfl{Our editing strategy operates within the generator's feature space to ensure seamless blending and preserve unedited regions. 
To achieve precise local control, we transform a user's 2D pixel-space edit into a precise mask in the parameterized UV space, where our features are defined.}
\kjh{
The user's edit is first localized by identifying the 2D edited regions via sketch differencing and dilation, yielding a 2D pixel mask $\mathcal{M}$. Next, we back-project this mask by casting rays through the edited regions to determine the set of 3D Gaussians contributing to those pixels. Following GaussianEditor~\cite{gaussianeditor}, the influence weight $w_i$ of each Gaussian is computed by accumulating its opacity--transmittance product over all masked pixels:
\begin{equation}
    w_i = \sum_{p \in \mathcal{M}} \alpha_i(p) \cdot T_i(p).
\end{equation}
We filter out back-face leakage by retaining only Gaussians with significant weights in the current view. These selected Gaussians are mapped to canonical UV space via FLAME coordinates to generate a precise binary mask $\mathbf{M}{\text{UV}}$. This mask is then resampled to match the resolution of each generator layer $k$, yielding layer-specific masks $\mathbf{M}{\text{UV}}^{(k)}$. Please refer to the supplementary material for evaluations of this strategy's robustness against extreme contour edits.
}

\noindent\textbf{Layer-by-Layer Feature Fusion.}
In contrast to artifact-prone methods that composite two sets of Gaussians directly in 3D space, we propose a layer-by-layer fusion strategy that operates within the StyleGAN generator's feature space. This approach achieves a natural blend by progressively guiding the synthesis process at multiple levels of abstraction.
The fusion is performed at every layer of the generator. Denote the intermediate feature map of the original (unedited) head as $\mathbf{f}_{k}^{\text{orig}}$ 
, and the corresponding feature map of the new (edited) head up to layer as $\mathbf{f}_{k}^{\text{new}}$.Instead of a simple blend, we use the features of the original head to guide the synthesis of the new one. Specifically, we create a fused feature map by selectively preserving the unedited regions from the original feature map:
\begin{equation}
\label{eq:fusion_step1}
\mathbf{f}_{k}^{\text{fused}} = (1 - \mathbf{M}_{\text{UV}}^{(k)}) \odot \mathbf{f}_{k}^{\text{orig}} + \mathbf{M}_{\text{UV}}^{(k)} \odot \mathbf{f}_{k}^{\text{new}},
\end{equation}
where $\odot$ denotes element-wise multiplication. 
 This fused tensor then serves as the input to the next synthesis layer of the generator, which in turn produces the next feature map:
 
\begin{equation}
\label{eq:fusion_step2}
\mathbf{f}_{k+1}^{\text{new}} = \text{Layer}_{k}(\mathbf{f}_{k}^{\text{fused}}, \mathcal{W}_{\text{new}}).
\end{equation}
where $\text{Layer}_{k}$ is the $k$-th layer in StyleGAN.
{By iteratively applying this process, unedited regions retain stable features while edited areas are progressively updated, yielding artifact-free and view-consistent results.}

\subsection{Training Strategy and Loss Functions}
\label{sec:training}

Our model is trained in three stages, each with a tailored objective: coarse generation, fine generation, and editing. During optimization, we employ a combination of pixel-wise (L1), perceptual, LPIPS, color-consistency, and adversarial losses to ensure geometric consistency, color controllability, and photorealistic details. Detailed training protocols, the usage of auxiliary decoders, and specific loss weights are provided in the supplementary material.

\section{Evaluation}
\label{sec:evaluation}


We conduct extensive experiments \lfl{(results in Sec.~\ref{sec:results}, comparisons in Sec.~\ref{sec:comparison}, ablations in Sec.~\ref{sec:ablation})} to evaluate SketchFaceGS in terms of quality, efficiency, and usability.



\noindent\textbf{Implementation Details.}
\lfl{We propose a stage-wise training strategy: the coarse stage is trained on a multi-view dataset synthesized from GGHead~\cite{gghead}, while the fine stage is trained on the single-view FFHQ dataset~\cite{Karras_2019}. Evaluation is performed on two test sets: (1) Generation set: 100 hand-drawn sketches collected from artists; (2) Editing set: 100 editing examples collected via our interactive system. For fair comparison, all baselines were re-trained or evaluated using their official codebases. For additional implementation details, please refer to the supplementary material.
}


\subsection{Results of Our Method}
\label{sec:results}

\begin{figure}[h!]
\centering
\includegraphics[width=\linewidth]{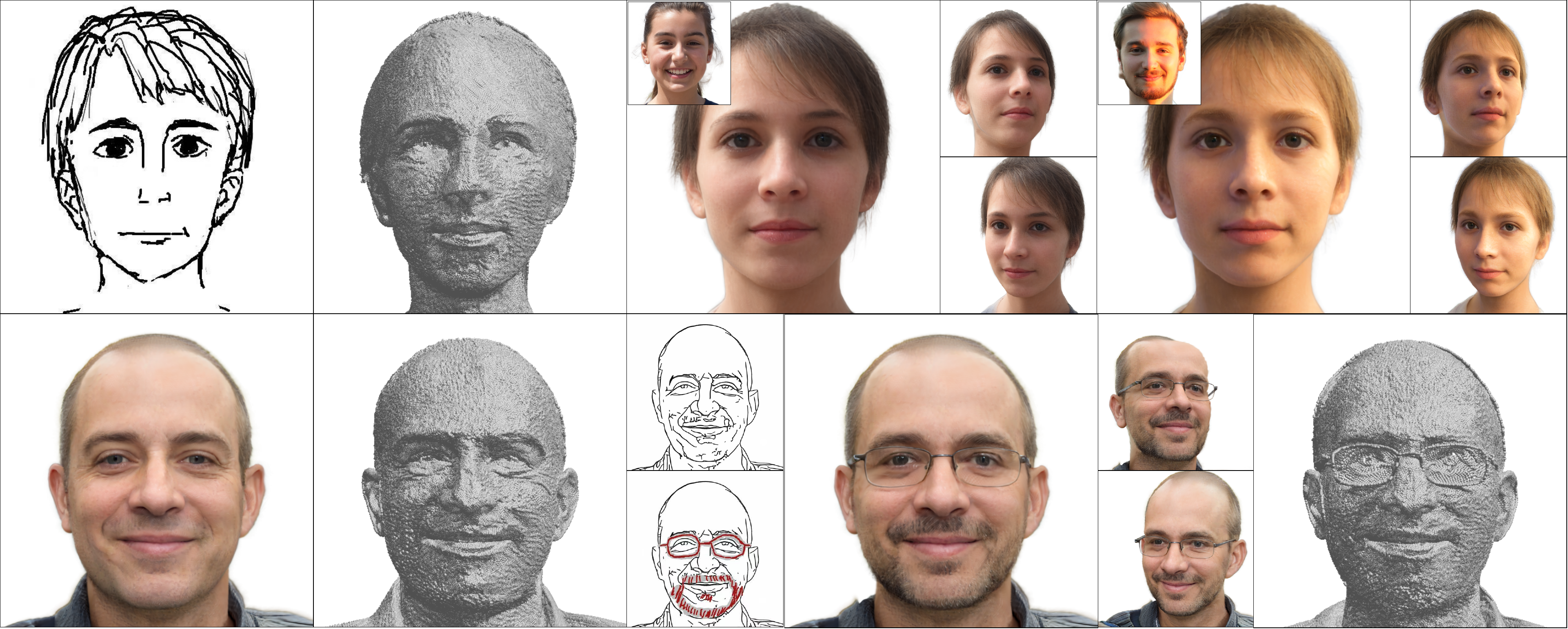}

\caption{
\lfl{Given a hand-drawn sketch and a reference image, our method produces a photorealistic 3D head (Top). Our method also support detailed local editing (Bottom).}
}
\label{fig:results} 
\end{figure}

\begin{figure}[t]
\centering
\includegraphics[width=\linewidth]{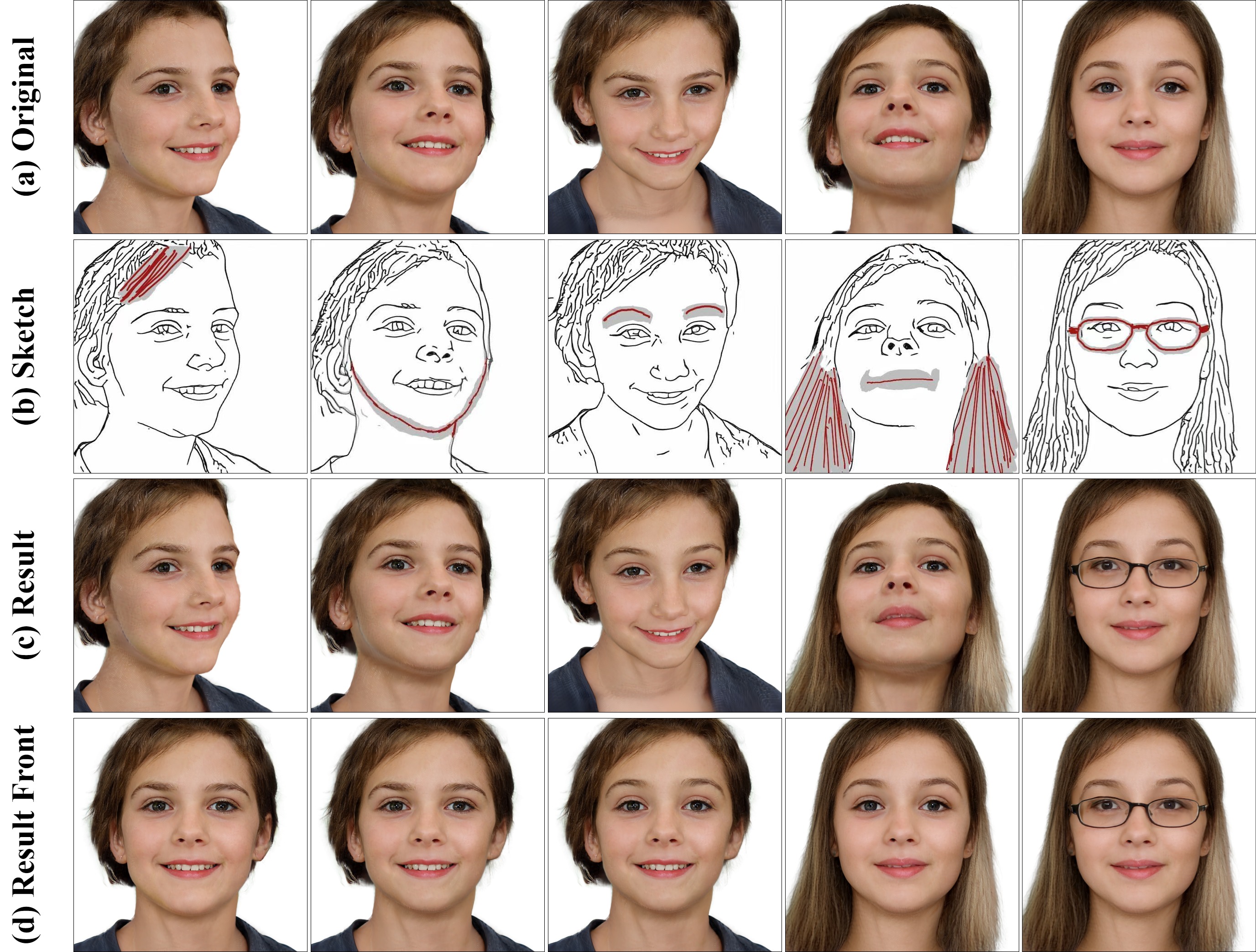}
\caption{\textbf{Results of continuous editing over five steps.} Progressively changing the sketches in different viewpoint achieves detailed and photorealistic editing results.}
\label{fig:multistep_editing}
\end{figure}

\lfl{
Figure~\ref{fig:results} (top) shows representative outputs synthesized from a single-view hand-drawn sketch with optional appearance references. Our pipeline reliably maps sparse stroke into a geometrically faithful 3DGS head. Silhouettes, hairlines, and stroke-level structural cues are preserved in geometry, while the final renderings exhibit high-frequency texture and consistent illumination across novel views.
\kjh{Because our model explicitly decouples geometry and appearance, the generated identity naturally emerges as a fusion of both modalities rather than a direct copy of the reference. Despite this structural adaptation to the sketch, key chromatic attributes (e.g., skin and hair color) are faithfully preserved.}
Figure~\ref{fig:results} (bottom) shows editing examples. Users can modify local regions, such as hair, glasses, and facial contours, from arbitrary viewpoints.
Edits accurately follow the input sketches, and unedited regions remain unchanged and blend naturally with the modified areas.

Our feed-forward approach is robust for multi-step, iterative editing. As shown in Fig.~\ref{fig:multistep_editing}, users can apply long sequences of diverse edits, such as altering hairstyle, facial contour, eyebrows, mouth, and eyes. Our UV Mask Fusion and layer-wise feature fusion ensure that each modification is precisely localized.
This stability makes our method exceptionally well-suited for practical, creative workflows that require continuous and repeated refinements.
}

\subsection{Comparison}
\label{sec:comparison}

\begin{figure*}[t]
\centering
\includegraphics[width=0.98\linewidth]{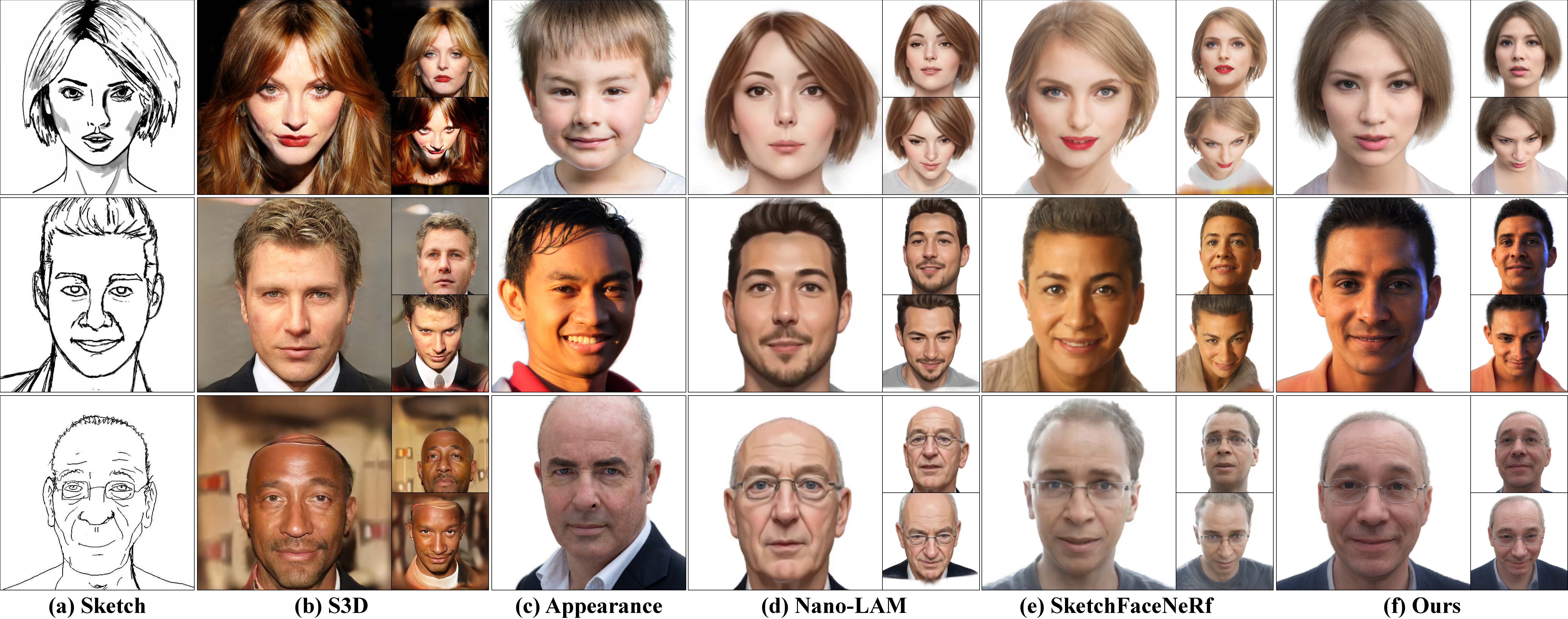}
\caption{\textbf{Qualitative comparison for sketch-to-3D generation.} In each example, (a) is the input sketch and (c) is the reference appearance. S3D (b) shows geometric inconsistencies. Nano-LAM (d) produces cartoonish results. SketchFaceNeRF (e) lacks fine-grained detail and realism. Our method (f) generates superior results with high fidelity to both geometry and appearance.}
\label{fig:comparison_generation}
\end{figure*}

\paragraph{Sketch-to-3D Head Generation.}
\kjh{To ensure a fair evaluation of the underlying architectures, we compare our method with leading sketch-to-3D approaches, including SketchFaceNeRF~\cite{gao2023sketchfacenerf} and S3D~\cite{song2025s3d}. Although we utilize GGHead while SketchFaceNeRF employs EG3D, the comparison remains valid as both generative priors are pre-trained on the same FFHQ dataset. We also include a two-stage baseline (Nano-LAM) that first generates a 2D image using Nano-Banana~\cite{google_gemini_2025} and subsequently reconstructs a 3DGS head using LAM~\cite{lam}.}
As shown in Fig.~\ref{fig:comparison_generation}, S3D struggles 
\lfl{with} hand-drawn sketches and produces geometric inconsistencies. 
\kjh{While utilizing prompt engineering on Nano-Banana~\cite{google_gemini_2025} can yield plausible 2D portraits, its reliance on LAM~\cite{lam} for single-view 3D lifting inevitably introduces geometric distortions and artifacts under large rotations.}
SketchFaceNeRF generates reasonable 3D shapes, but its coarse tri-plane prediction 
often fails to recover fine-grained details from the input sketches (e.g., complex hair strands in the 2nd row). Its outputs also fall short of photorealistic fidelity.
\lfl{By directly synthesizing Gaussian attributes through a StyleGAN-based UV manifold representation, we recover sharper geometry and clearer details while remaining faithful to the input sketch. The quantitative evaluation is reported in Table~\ref{tab:generation_metrics}.}
Our method achieves the best FID and KID on the held-out sketch set, demonstrating improved realism and sketch faithfulness.

\begin{table}[h!]
\centering
\caption{\textbf{Quantitative comparison for sketch-based 3D head generation.} Lower values indicate better performance.}
\label{tab:generation_metrics}
\begin{tabular}{lcc}
\toprule
Method & FID $\downarrow$ & KID ($\times 100$) $\downarrow$ \\
\midrule
S3D~\cite{song2025s3d} & 96.03 & 4.50 $\pm$ 1.0 \\
Nano-LAM~\cite{google_gemini_2025, lam} & 133.72 & 7.61 $\pm$ 0.9 \\
SketchFaceNeRF~\cite{gao2023sketchfacenerf} & 94.94 & 4.53 $\pm$ 0.6 \\
Ours & \textbf{92.65} & \textbf{4.00 $\pm$ 0.4} \\
\bottomrule
\end{tabular}
\end{table}

\paragraph{Sketch-based 3D Head Editing.}

\begin{figure*}[t]
\centering
\includegraphics[width=0.98\linewidth]{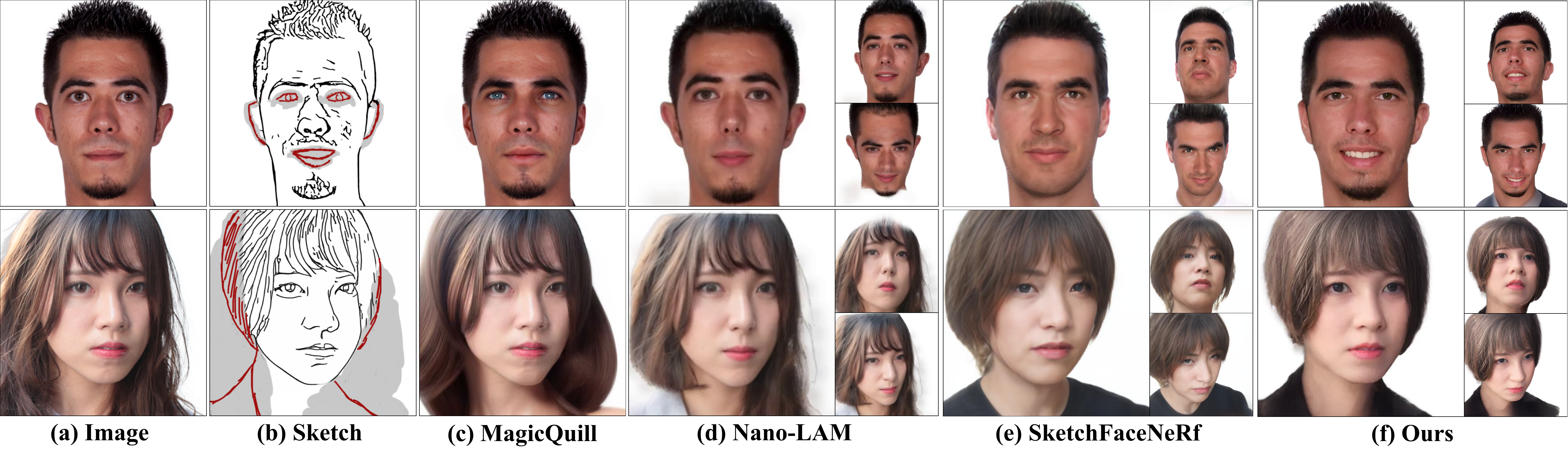}
\caption{\textbf{Qualitative comparison for sketch-based 3D head editing.} Given an original head (a) and an edit sketch (b), MagicQuill (c) produces stylized/blurry results; Nano-LAM (d) suffers from imprecise editing with poor novel-view quality. SketchFaceNeRF (e) is optimization-based and non-interactive. Our method (f) achieves real-time, high-quality editing that faithfully follows the sketch.}
\label{fig:comparison_editing}
\end{figure*}

\lfl{For editing, we compare against SketchFaceNeRF~\cite{gao2023sketchfacenerf}, MagicQuill~\cite{liu2025magicquill}, and the similar Nano-LAM 2D-to-3D editing baseline. MagicQuill, a 2D ControlNet-based method, is evaluated from the original rendering viewpoint for fairness. As shown in Fig.~\ref{fig:comparison_editing}, MagicQuill can perform edits but often introduces stylization and identity shifts.
\kjh{Although Nano-Banana~\cite{google_gemini_2025} can perform 2D edits in the image space, relying on LAM~\cite{lam} to lift these modifications to 3D often results in poor multi-view consistency. Furthermore, the generation process of Nano-Banana takes over 10 seconds, which entirely precludes the real-time interactivity achieved by our feed-forward architecture.}
SketchFaceNeRF produces photorealistic results but relies on slow optimization that lacks precision for fine stroke intent.
\kjh{Our method achieves the most accurate sketch-aligned edits while inherently preserving the structural integrity of the original face. As demonstrated in Table~\ref{tab:unedited_metrics}, our approach shows superior identity preservation in non-edited regions compared to SketchFaceNeRF (both before and after its optimization), avoiding the identity drift common in optimization-based methods.}
It is also substantially faster: as reported in Table~\ref{tab:editing_metrics}, we achieve the best FID/KID, an end-to-end edit latency of ~0.3 s, and up to 243 FPS rendering. This enables our approach to achieve practical, interactive editing workflows.
Notably, SketchFaceNeRF requires \textbf{10 s} per edit.}

\begin{table}[h!] \footnotesize
\centering
\caption{\textbf{Quantitative comparison for sketch-based 3D head editing.} We report image quality (FID, KID) and interaction/rendering performance. 
}
\label{tab:editing_metrics}
\begin{tabular}{lcccc}
\toprule
Method & FID $\downarrow$ & KID ($\times 100$) $\downarrow$ & Time (s) $\downarrow$ & FPS $\uparrow$ \\
\midrule
MagicQuill & 46.48 & 0.78 $\pm$ 0.2 & $\sim$6.0 & --- \\
Nano-LAM & 74.26 & 3.01 $\pm$ 0.3 & $\sim$15.0 & \textbf{281} \\
SketchFaceNeRF &  62.49 & 2.65 $\pm$ 0.3 & $\sim$10.0 & 42 \\
Ours & \textbf{44.60} & \textbf{0.69 $\pm$ 0.2} & \textbf{$\sim$0.3} & 243 \\
\bottomrule
\end{tabular}
\end{table}

\begin{table}[h!] \footnotesize
\centering
\caption{\textbf{Quantitative comparison of identity preservation in unedited regions with SketchFaceNeRF.} Higher values indicate better preservation of the original content outside the edited area.}
\label{tab:unedited_metrics}
\begin{tabular}{lccc}
\toprule
Metric & SF-NeRF (w/o opt.) & SF-NeRF (opt.) & Ours \\
\midrule
PSNR $\uparrow$ & 22.30 & 27.78 & \textbf{31.12} \\
SSIM $\uparrow$ & 0.90 & 0.95 & \textbf{0.97} \\
\bottomrule
\end{tabular}
\end{table}
\subsection{Ablation Study}
\label{sec:ablation}

\begin{figure*}[t]
\centering
\includegraphics[width=0.98\linewidth]{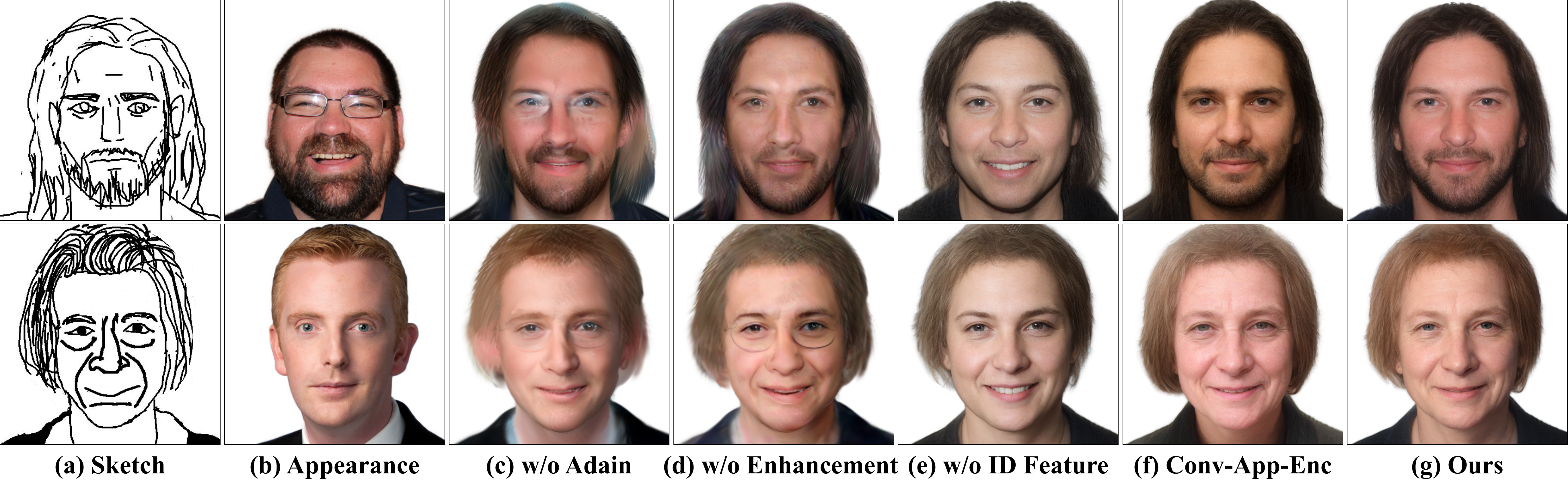}
\caption{\textbf{Ablation study on the generation pipeline.} (a) Input sketch. (b) Input appearance. (c) Without AdaIN alignment, identity conflicts arise. (d) Without the enhancement module, details are lost. (e) Without identity vectors, identity is inconsistent. (f) With a simple CNN for appearance, style transfer fails. (g) Our full model yields the best result.}
\label{fig:ablation_generation}
\end{figure*}

\begin{figure}[t]
\centering
\includegraphics[width=\linewidth]{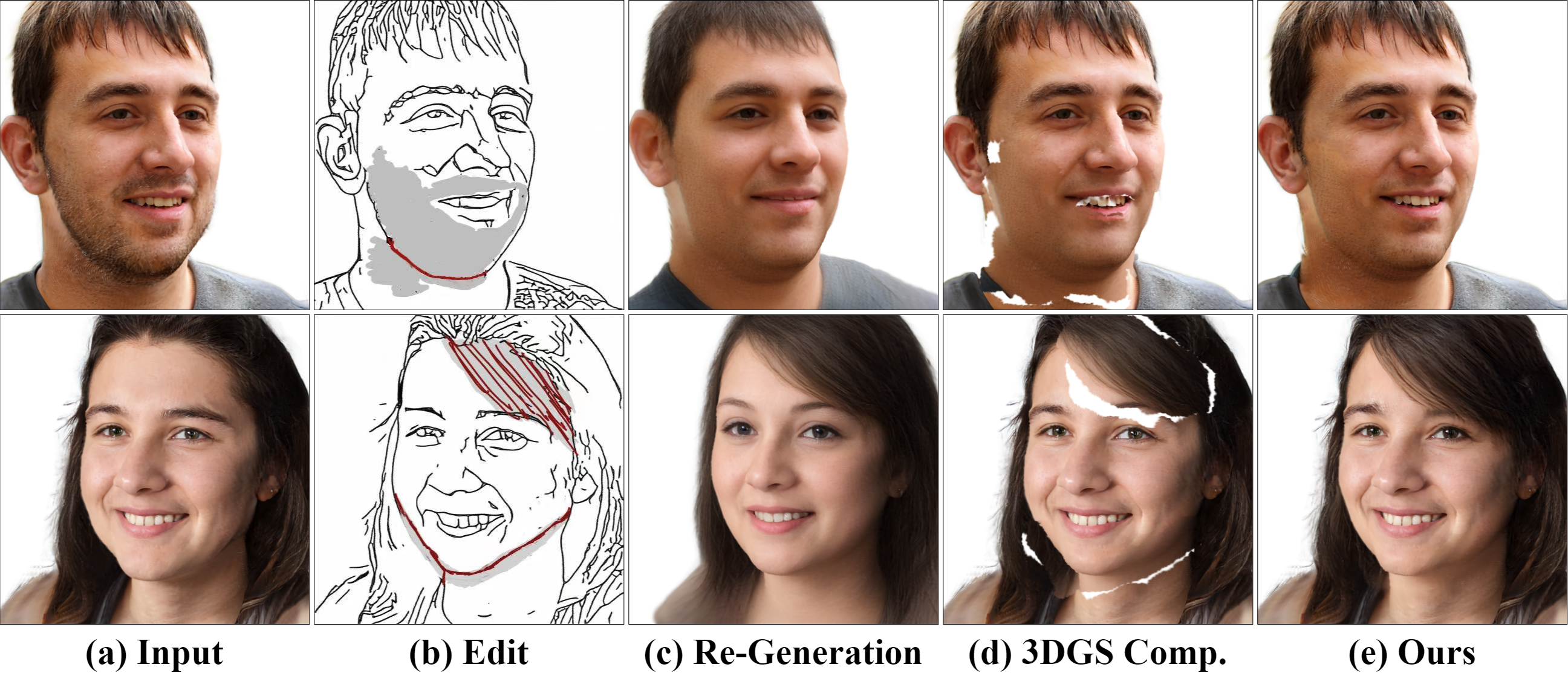}
\caption{
\textbf{Ablation study on the editing module.} \lfl{Our method (e) achieves the superior results compared with re-generation (c) and direct 3DGS composition (d).}
}
\label{fig:ablation_editing}
\end{figure}


\paragraph{Generation Pipeline Ablations.}
We examine the following settings: (a) \emph{w/o enhancement module} --- removing the UV Feature Enhancement stage; (b) \emph{w/o translation network} --- disabling the AdaIN-based alignment between geometry and appearance; (c) \emph{conv appearance encoder} --- replacing the appearance Transformer with a simple CNN encoder; and (d) \emph{w/o global ID feature} --- removing the global identity vectors.
As shown in Fig.~\ref{fig:ablation_generation}, \lfl{removing the} enhancement module (a) \lfl{keeps coarse geometry but yields oversmoothed results without realistic details.} 
Disabling the translation network (b) causes severe artifacts when sketch and reference identities conflict. 
\lfl{Using a CNN encoder} (c) weakens style transfer and causes color mismatches. Removing global identity vectors (d) degrades identity-specific attributes. 
\lfl{Table~\ref{tab:ablation_generation} shows that the full model consistently outperforms all variants.}

\begin{table}[h!]
\centering
\caption{\textbf{Quantitative ablation study for the generation pipeline.} All ablated versions show a significant drop in performance compared to our full model. Lower values are better.}
\label{tab:ablation_generation}
\begin{tabular}{lcc}
\toprule
Generation Ablations & FID $\downarrow$ & KID ($\times 100$) $\downarrow$ \\
\midrule
Full Model (Ours) & \textbf{92.65} & \textbf{4.00} $\pm$ 0.4\\
\midrule
w/o Enhancement Module & 104.08 & 6.14 $\pm$ 1.3 \\
w/o Translation Network & 108.26 & 8.10 $\pm$ 0.7\\
Conv Appearance Encoder & 97.87 & 5.77 $\pm$ 0.5 \\
w/o Global ID Feature & 98.73 & 5.74 $\pm$ 0.6 \\
\bottomrule
\end{tabular}
\end{table}

\paragraph{Editing Module Ablations.}
We compare our layer-wise feature fusion with two alternatives: (c) \emph{direct regeneration} --- regenerating \lfl{the entire} 
head from the edit sketch without fusion, and (d) \emph{3D Gaussian compositing} --- directly replacing affected Gaussians in 3D space. As shown in Fig.~\ref{fig:ablation_editing}, direct regeneration (c) discards unedited context and fails to preserve identity. 3D Gaussian compositing (d) introduces conspicuous seams and mismatches at edit boundaries. In contrast, our layer-wise feature fusion produces visually coherent, artifact-free transitions by blending at multiple feature resolutions. The superiority of our approach is quantitatively validated in Table~\ref{tab:ablation_editing}, where our method achieves substantially better FID and KID scores.

\begin{table}[h!]
\centering
\caption{\textbf{Quantitative ablation study for the editing module.} Our layer-wise feature fusion significantly outperforms alternative strategies. Lower values are better.}
\label{tab:ablation_editing}
\begin{tabular}{lcc}
\toprule
Editing Ablations & FID $\downarrow$ & KID ($\times 100$) $\downarrow$ \\
\midrule
Full Model (Ours) & \textbf{44.60} & \textbf{0.69} $\pm$ 0.2 \\
\midrule
Re-Generation & 88.15 & 3.35 $\pm$ 0.7 \\
3D Gaussian Compositing & 68.42 & 1.89 $\pm$ 0.2\\
\bottomrule
\end{tabular}
\end{table}


\subsection{Applications}
\label{sec:applications}

Beyond the core generation and editing tasks, our method also enables practical downstream applications such as local appearance transfer and fast editable inversion of real facial images. Please refer to the supplementary material for detailed demonstrations and visual results.



\section{Conclusions and Discussions}
\label{sec:conclusion}



\lfl{
This paper presents SketchFaceGS, a real-time 3DGS-based framework for human head generation and editing to enable interactive avatar creation.
For generation, we adopt a coarse-to-fine generation paradigm.
In the coarse stage, a dual-transformer feedforward network extracts features from sketch and appearance images respectively, following by fusion network to predicts a coarse UV map.
In the fine stage, we translate this UV map into a global latent and modulation parameters to generate high-quality 3D faces based on a pretrained 3D GAN.
For editing, we introduce a UV Mask Fusion strategy to ensure effective local edits while preserving unedited regions. }
\kjh{
\paragraph{Limitations and Future Work.}
Despite its advantages, our method has limitations pointing to future research. First, geometric sketch-reference discrepancies can cause slight identity shifts despite chromatic preservation; we aim to mitigate this with identity consistency losses or advanced encoders. Second, GGHead prior reliance limits handling of rare accessories, extreme occlusions, and OOD inputs; expanding data or adding specialized accessory modules could resolve this. Finally, our static head editing framework will extend to facial animation, leveraging UV-masked fusion's motion control compatibility with a robust drivable 3DGS-GAN backbone.}
\section*{Acknowledgments}
This work was supported by the National Natural Science Foundation of China (No. 62472205, 62322210, 62561160115, U25A20480, U2441241), the Innovation Funding of ICT, CAS (No. E561100), Academic Leaders Training Program of Jiangxi Province (20232BCJ22001), Key Project of Jiangxi Natural Science Foundation (20224ACB202008), Key R\&D Plan of Jiangxi Province (20232BBE50022), and the HNXJ Philanthropy Foundation (No. KY24010).
{
    \small
    \bibliographystyle{ieeenat_fullname}
    \bibliography{main}

@String(CVPR= {IEEE Conf. Comput. Vis. Pattern Recog.})

@String(ICCV= {Int. Conf. Comput. Vis.})

@String(ECCV= {Eur. Conf. Comput. Vis.})

@String(NIPS= {Adv. Neural Inform. Process. Syst.})

@String(TOG= TOG)

@String(TVCG  = {IEEE Trans. Vis. Comput. Graph.})

@String(TMM  = {IEEE Trans. Multimedia})

@String(ACMMM= {ACM Int. Conf. Multimedia})

@String(ICLR = {Int. Conf. Learn. Represent.})

@String(SIGGRAPH = {Proc. ACM SIGGRAPH})

@String(CVPR  = {CVPR})

@String(ICCV  = {ICCV})

@String(ECCV  = {ECCV})

@String(NIPS  = {NeurIPS})

@String(TOG   = {ACM TOG})

@String(TVCG  = {IEEE TVCG})

@String(TMM   =	{IEEE TMM})

@String(ACMMM = {ACM MM})

@String(ICLR  = {ICLR})

@article{Portenier2018,
  author       = {Tiziano Portenier and
                  Qiyang Hu and
                  Attila Szab{\'{o}} and
                  Siavash Arjomand Bigdeli and
                  Paolo Favaro and
                  Matthias Zwicker},
  title        = {Faceshop: deep sketch-based face image editing},
  journal      = TOG,
  volume       = {37},
  number       = {4},
  pages        = {99},
  year         = {2018},
  url          = {https://doi.org/10.1145/3197517.3201393},
  doi          = {10.1145/3197517.3201393},
  bibsource    = {dblp computer science bibliography, https://dblp.org}
}

@misc{Maya2019,
  author = {{Autodesk, Inc.}},
  title = {Autodesk {Maya}},
  year = {2019},
  howpublished = {\url{https://www.autodesk.com/maya}},
}

@article{Bergman,
  title={Generative neural articulated radiance fields},
  author={Bergman, Alexander and Kellnhofer, Petr and Yifan, Wang and Chan, Eric and Lindell, David and Wetzstein, Gordon},
  journal=NIPS,
  volume={35},
  pages={19900--19916},
  year={2022}
}

@inproceedings{Gu2021,
  author       = {Jiatao Gu and
                  Lingjie Liu and
                  Peng Wang and
                  Christian Theobalt},
  title        = {{StyleNeRF}: {A} Style-based {3D} Aware Generator for High-resolution
                  Image Synthesis},
  booktitle    = ICLR,
  publisher    = {OpenReview.net},
  year         = {2022},
  url          = {https://openreview.net/forum?id=iUuzzTMUw9K},
  bibsource    = {dblp computer science bibliography, https://dblp.org}
}

@article{Schwarz2020,   
  title={Graf: Generative radiance fields for 3d-aware image synthesis},
  author={Schwarz, Katja and Liao, Yiyi and Niemeyer, Michael and Geiger, Andreas},
  journal=NIPS,
  volume={33},
  pages={20154--20166},
  year={2020}
}

@article{Mildenhall2020,
  title={Nerf: Representing scenes as neural radiance fields for view synthesis},
  author={Mildenhall, Ben and Srinivasan, Pratul P and Tancik, Matthew and Barron, Jonathan T and Ramamoorthi, Ravi and Ng, Ren},
  journal={Communications of the ACM},
  volume={65},
  number={1},
  pages={99--106},
  year={2021},
  publisher={ACM New York, NY, USA}
}

@article{gaussian_splatting,
  author       = {Bernhard Kerbl and
                  Georgios Kopanas and
                  Thomas Leimk{\"{u}}hler and
                  George Drettakis},
  title        = {{3D} Gaussian Splatting for Real-Time Radiance Field Rendering},
  journal      = TOG,
  volume       = {42},
  number       = {4},
  pages        = {139:1--139:14},
  year         = {2023},
  url          = {https://doi.org/10.1145/3592433},
  doi          = {10.1145/3592433},
  bibsource    = {dblp computer science bibliography, https://dblp.org}
}

@inproceedings{gghead,
  title={Gghead: Fast and generalizable 3d gaussian heads},
  author={Kirschstein, Tobias and Giebenhain, Simon and Tang, Jiapeng and Georgopoulos, Markos and Nie{\ss}ner, Matthias},
  booktitle=SIGGRAPH,
  pages={1--11},
  year={2024}
}

@article{Hierarchical_gan,
  title={Gsgan: Adversarial learning for hierarchical generation of 3d gaussian splats},
  author={Hyun, Sangeek and Heo, Jae-Pil},
  journal=NIPS,
  volume={37},
  pages={67987--68012},
  year={2024}
}

@inproceedings{wu2024,
  title={Gaussctrl: Multi-view consistent text-driven 3d gaussian splatting editing},
  author={Wu, Jing and Bian, Jia-Wang and Li, Xinghui and Wang, Guangrun and Reid, Ian and Torr, Philip and Prisacariu, Victor Adrian},
  booktitle=ECCV,
  pages={55--71},
  year={2024},
  organization={Springer}
}

@inproceedings{gaussianeditor,
  title={Gaussianeditor: Editing 3d gaussians delicately with text instructions},
  author={Wang, Junjie and Fang, Jiemin and Zhang, Xiaopeng and Xie, Lingxi and Tian, Qi},
  booktitle=CVPR,
  pages={20902--20911},
  year={2024}
}

@misc{vachha2024,
         author = {Vachha, Cyrus and Haque, Ayaan},
         title = {Instruct-GS2GS: Editing 3D Gaussian Splats with Instructions},
         year = {2024},
         url = {https://instruct-gs2gs.github.io/}
        }

@article{Han_2018, 
  title={Caricatureshop: Personalized and photorealistic caricature sketching},
  author={Han, Xiaoguang and Hou, Kangcheng and Du, Dong and Qiu, Yuda and Cui, Shuguang and Zhou, Kun and Yu, Yizhou},
  journal=TVCG,
  volume={26},
  number={7},
  pages={2349--2361},
  year={2018},
  publisher={IEEE}
}

@article{chen2020deep,
  author       = {Shu{-}Yu Chen and
                  Wanchao Su and
                  Lin Gao and
                  Shihong Xia and
                  Hongbo Fu},
  title        = {DeepFaceDrawing: deep generation of face images from sketches},
  journal      = TOG,
  volume       = {39},
  number       = {4},
  pages        = {72},
  year         = {2020},
  url          = {https://doi.org/10.1145/3386569.3392386},
  doi          = {10.1145/3386569.3392386},
  bibsource    = {dblp computer science bibliography, https://dblp.org}
}

@article{chen2021deepfaceediting,
author = {Chen, Shu-Yu and Liu, Feng-Lin and Lai, Yu-Kun and Rosin, Paul L. and Li, Chunpeng and Fu, Hongbo and Gao, Lin},
title = {DeepFaceEditing: deep face generation and editing with disentangled geometry and appearance control},
year = {2021},
issue_date = {August 2021},
publisher = {Association for Computing Machinery},
address = {New York, NY, USA},
volume = {40},
number = {4},
issn = {0730-0301},
url = {https://doi.org/10.1145/3450626.3459760},
doi = {10.1145/3450626.3459760},
journal = TOG,
month = jul,
articleno = {90},
numpages = {15}
}

@inproceedings{Jo_Park_2019,   
  title={{SC-FEGAN}: Face editing generative adversarial network with user's sketch and color},
  author={Jo, Youngjoo and Park, Jongyoul},
  booktitle=ICCV,
  pages={1745--1753},
  year={2019}
}

@article{gao2023sketchfacenerf,
  title={{SketchFaceNeRF}: Sketch-based facial generation and editing in neural radiance fields},
  author={Gao, Lin and Liu, Feng-Lin and Chen, Shu-Yu and Jiang, Kaiwen and Li, Chunpeng and Lai, Yukun and Fu, Hongbo},
  journal=TOG,
  volume={42},
  number={4},
  year={2023},
  publisher={Association for Computing Machinery}
}

@inproceedings{lam,
  title={{LAM}: large avatar model for one-shot animatable gaussian head},
  author={He, Yisheng and Gu, Xiaodong and Ye, Xiaodan and Xu, Chao and Zhao, Zhengyi and Dong, Yuan and Yuan, Weihao and Dong, Zilong and Bo, Liefeng},
  booktitle=SIGGRAPH,
  pages={1--13},
  year={2025}
}

@inproceedings{gfp-gan,  
  title={Towards real-world blind face restoration with generative facial prior},
  author={Wang, Xintao and Li, Yu and Zhang, Honglun and Shan, Ying},
  booktitle=CVPR,
  pages={9168--9178},
  year={2021}
}

@article{gan,
  title={Generative adversarial nets},
  author={Goodfellow, Ian J and Pouget-Abadie, Jean and Mirza, Mehdi and Xu, Bing and Warde-Farley, David and Ozair, Sherjil and Courville, Aaron and Bengio, Yoshua},
  journal=NIPS,
  volume={27},
  year={2014}
}

@inproceedings{eg3d,    title={Efficient geometry-aware 3d generative adversarial networks},
  author={Chan, Eric R and Lin, Connor Z and Chan, Matthew A and Nagano, Koki and Pan, Boxiao and De Mello, Shalini and Gallo, Orazio and Guibas, Leonidas J and Tremblay, Jonathan and Khamis, Sameh and others},
  booktitle=CVPR,
  pages={16123--16133},
  year={2022}
}

@article{ho2020denoising,
  title={Denoising diffusion probabilistic models},
  author={Ho, Jonathan and Jain, Ajay and Abbeel, Pieter},
  journal=NIPS,
  volume={33},
  pages={6840--6851},
  year={2020}
}

@inproceedings{Chen_Zhang_2019,    title={Learning implicit fields for generative shape modeling},
  author={Chen, Zhiqin and Zhang, Hao},
  booktitle=CVPR,
  pages={5939--5948},
  year={2019}
}

@inproceedings{Kanazawa_2018,
  title={Learning category-specific mesh reconstruction from image collections},
  author={Kanazawa, Angjoo and Tulsiani, Shubham and Efros, Alexei A and Malik, Jitendra},
  booktitle=ECCV,
  pages={371--386},
  year={2018}
}

@inproceedings{Gadelha_2016,
  title={3d shape induction from 2d views of multiple objects},
  author={Gadelha, Matheus and Maji, Subhransu and Wang, Rui},
  booktitle={2017 international conference on 3d vision (3DV)},
  pages={402--411},
  year={2017},
  organization={IEEE}
}

@inproceedings{Henzler_2019,  
  title={Escaping plato's cave: 3d shape from adversarial rendering},
  author={Henzler, Philipp and Mitra, Niloy J and Ritschel, Tobias},
  booktitle=ICCV,
  pages={9984--9993},
  year={2019}
}

@inproceedings{2021pi,
  title={pi-gan: Periodic implicit generative adversarial networks for 3d-aware image synthesis},
  author={Chan, Eric R and Monteiro, Marco and Kellnhofer, Petr and Wu, Jiajun and Wetzstein, Gordon},
  booktitle=CVPR,
  pages={5799--5809},
  year={2021}
}

@inproceedings{Nguyen-Phuoc_2019,  
  title={Hologan: Unsupervised learning of 3d representations from natural images},
  author={Nguyen-Phuoc, Thu and Li, Chuan and Theis, Lucas and Richardt, Christian and Yang, Yong-Liang},
  booktitle=ICCV,
  pages={7588--7597},
  year={2019}
}

@inproceedings{Niemeyer_2021,   
  title={Giraffe: Representing scenes as compositional generative neural feature fields},
  author={Niemeyer, Michael and Geiger, Andreas},
  booktitle=CVPR,
  pages={11453--11464},
  year={2021}
}

@inproceedings{Xue,
  title={Giraffe hd: A high-resolution 3d-aware generative model},
  author={Xue, Yang and Li, Yuheng and Singh, Krishna Kumar and Lee, Yong Jae},
  booktitle=CVPR,
  pages={18440--18449},
  year={2022}
}

@inproceedings{Karras_2019,  
  title={A style-based generator architecture for generative adversarial networks},
  author={Karras, Tero and Laine, Samuli and Aila, Timo},
  booktitle=CVPR,
  pages={4401--4410},
  year={2019}
}

@inproceedings{OrEl_2022, 
  title={Stylesdf: High-resolution 3d-consistent image and geometry generation},
  author={Or-El, Roy and Luo, Xuan and Shan, Mengyi and Shechtman, Eli and Park, Jeong Joon and Kemelmacher-Shlizerman, Ira},
  booktitle=CVPR,
  pages={13503--13513},
  year={2022}
}

@article{song2025s3d,
  title={S3d: Sketch-driven 3d model generation},
  author={Song, Hail and Shin, Wonsik and Lee, Naeun and Chung, Soomin and Kwak, Nojun and Woo, Woontack},
  journal={arXiv preprint arXiv:2505.04185},
  year={2025}
}

@inproceedings{stylegan2,
  title={Analyzing and improving the image quality of stylegan},
  author={Karras, Tero and Laine, Samuli and Aittala, Miika and Hellsten, Janne and Lehtinen, Jaakko and Aila, Timo},
  booktitle=CVPR,
  pages={8110--8119},
  year={2020}
}

@article{oquab2023dinov2,
  author       = {Maxime Oquab and
                  Timoth{\'{e}}e Darcet and
                  Th{\'{e}}o Moutakanni and
                  Huy V. Vo and
                  Marc Szafraniec and
                  Vasil Khalidov and
                  Pierre Fernandez and
                  Daniel Haziza and
                  Francisco Massa and
                  Alaaeldin El{-}Nouby and
                  Mido Assran and
                  Nicolas Ballas and
                  Wojciech Galuba and
                  Russell Howes and
                  Po{-}Yao Huang and
                  Shang{-}Wen Li and
                  Ishan Misra and
                  Michael Rabbat and
                  Vasu Sharma and
                  Gabriel Synnaeve and
                  Hu Xu and
                  Herv{\'{e}} J{\'{e}}gou and
                  Julien Mairal and
                  Patrick Labatut and
                  Armand Joulin and
                  Piotr Bojanowski},
  title        = {{DINOv2}: Learning Robust Visual Features without Supervision},
  journal      = {Trans. Mach. Learn. Res.},
  volume       = {2024},
  year         = {2024},
  url          = {https://openreview.net/forum?id=a68SUt6zFt},
  bibsource    = {dblp computer science bibliography, https://dblp.org}
}

@inproceedings{LinesToFacePhoto,
  author       = {Yuhang Li and
                  Xuejin Chen and
                  Feng Wu and
                  Zheng{-}Jun Zha},
  editor       = {Laurent Amsaleg and
                  Benoit Huet and
                  Martha A. Larson and
                  Guillaume Gravier and
                  Hayley Hung and
                  Chong{-}Wah Ngo and
                  Wei Tsang Ooi},
  title        = {LinesToFacePhoto: Face Photo Generation From Lines With Conditional
                  Self-Attention Generative Adversarial Networks},
  booktitle    = ACMMM,
  publisher    = {{ACM}},
  year         = {2019},
  url          = {https://doi.org/10.1145/3343031.3350854},
  doi          = {10.1145/3343031.3350854},
  bibsource    = {dblp computer science bibliography, https://dblp.org}
}

@inproceedings{DeepFacePencil,
  author       = {Yuhang Li and
                  Xuejin Chen and
                  Binxin Yang and
                  Zihan Chen and
                  Zhihua Cheng and
                  Zheng{-}Jun Zha},
  editor       = {Chang Wen Chen and
                  Rita Cucchiara and
                  Xian{-}Sheng Hua and
                  Guo{-}Jun Qi and
                  Elisa Ricci and
                  Zhengyou Zhang and
                  Roger Zimmermann},
  title        = {DeepFacePencil: Creating Face Images from Freehand Sketches},
  booktitle    = ACMMM,
  pages        = {991--999},
  publisher    = {{ACM}},
  year         = {2020},
  url          = {https://doi.org/10.1145/3394171.3413684},
  doi          = {10.1145/3394171.3413684},
  bibsource    = {dblp computer science bibliography, https://dblp.org}
}

@inproceedings{ControllableFace,
  title={Controllable face sketch-photo synthesis with flexible generative priors},
  author={Cheng, Kun and Zhu, Mingrui and Wang, Nannan and Li, Guozhang and Wang, Xiaoyu and Gao, Xinbo},
  booktitle=ACMMM,
  pages={6959--6968},
  year={2023}
}

@article{DeepSketch2Face,
  title={DeepSketch2Face: a deep learning based sketching system for {3D} face and caricature modeling},
  author={Han, Xiaoguang and Gao, Chang and Yu, Yizhou},
  journal=TOG,
  volume={36},
  number={4},
  pages={1--12},
  year={2017},
  publisher={ACM New York, NY, USA}
}

@article{yang2021learning,
  title={Learning {3D} face reconstruction from a single sketch},
  author={Yang, Li and Wu, Jing and Huo, Jing and Lai, Yu-Kun and Gao, Yang},
  journal={Graphical Models},
  volume={115},
  pages={101102},
  year={2021},
  publisher={Elsevier}
}

@inproceedings{yang2020deep,
  title={Deep plastic surgery: Robust and controllable image editing with human-drawn sketches},
  author={Yang, Shuai and Wang, Zhangyang and Liu, Jiaying and Guo, Zongming},
  booktitle=ECCV,
  pages={601--617},
  year={2020},
  organization={Springer}
}

@inproceedings{zeng2022sketchedit,
  title={Sketchedit: Mask-free local image manipulation with partial sketches},
  author={Zeng, Yu and Lin, Zhe and Patel, Vishal M},
  booktitle=CVPR,
  pages={5951--5961},
  year={2022}
}

@article{du2020sanihead,
  title={SAniHead: Sketching animal-like {3D} character heads using a view-surface collaborative mesh generative network},
  author={Du, Dong and Han, Xiaoguang and Fu, Hongbo and Wu, Feiyang and Yu, Yizhou and Cui, Shuguang and Liu, Ligang},
  journal=TVCG,
  volume={28},
  number={6},
  pages={2415--2429},
  year={2020},
  publisher={IEEE}
}

@inproceedings{luo2021simpmodeling,
  title={Simpmodeling: Sketching implicit field to guide mesh modeling for 3d animalmorphic head design},
  author={Luo, Zhongjin and Zhou, Jie and Zhu, Heming and Du, Dong and Han, Xiaoguang and Fu, Hongbo},
  booktitle={The 34th annual ACM symposium on user interface software and technology},
  pages={854--863},
  year={2021}
}

@inproceedings{liu2025magicquill,
  title={Magicquill: An intelligent interactive image editing system},
  author={Liu, Zichen and Yu, Yue and Ouyang, Hao and Wang, Qiuyu and Cheng, Ka Leong and Wang, Wen and Liu, Zhiheng and Chen, Qifeng and Shen, Yujun},
  booktitle=CVPR,
  pages={13072--13082},
  year={2025}
}

@inproceedings{kangle2023pix2pix3d,
  title={3d-aware conditional image synthesis},
  author={Deng, Kangle and Yang, Gengshan and Ramanan, Deva and Zhu, Jun-Yan},
  booktitle=CVPR,
  pages={4434--4445},
  year={2023}
}

@inproceedings{hong2023lrm,
  author       = {Yicong Hong and
                  Kai Zhang and
                  Jiuxiang Gu and
                  Sai Bi and
                  Yang Zhou and
                  Difan Liu and
                  Feng Liu and
                  Kalyan Sunkavalli and
                  Trung Bui and
                  Hao Tan},
  title        = {{LRM:} Large Reconstruction Model for Single Image to {3D}},
  booktitle    = ICLR,
  publisher    = {OpenReview.net},
  year         = {2024},
  url          = {https://openreview.net/forum?id=sllU8vvsFF},
  bibsource    = {dblp computer science bibliography, https://dblp.org}
}

@inproceedings{weng2024template,
  title={Template-Free Single-View {3D} Human Digitalization with Diffusion-Guided LRM},
  author={Zhenzhen Weng and Jingyuan Liu and Hao Tan and Zhan Xu and Yang Zhou and Serena Yeung-Levy and Jimei Yang},
  year={2024},
  url={https://api.semanticscholar.org/CorpusID:267069397}
}

@article{pan2024humansplat,
  title={Humansplat: Generalizable single-image human gaussian splatting with structure priors},
  author={Pan, Panwang and Su, Zhuo and Lin, Chenguo and Fan, Zhen and Zhang, Yongjie and Li, Zeming and Shen, Tingting and Mu, Yadong and Liu, Yebin},
  journal=NIPS,
  volume={37},
  pages={74383--74410},
  year={2024}
}

@article{chu2024gagavatar,
  title={Generalizable and animatable gaussian head avatar},
  author={Chu, Xuangeng and Harada, Tatsuya},
  journal=NIPS,
  volume={37},
  pages={57642--57670},
  year={2024}
}

@misc{google_gemini_2025,
  author = {Fortin, Alisa and Vernade, Guillaume and Kampf, Kat and Reshi, Ammaar},
  title = {Introducing {Gemini} 2.5 {Flash Image}, our state-of-the-art image model},
  year = {2025},
  howpublished = {\url{https://developers.googleblog.com/introducing-gemini-2-5-flash-image/}},
}

@inproceedings{Perceptual,
  title={Perceptual losses for real-time style transfer and super-resolution},
  author={Johnson, Justin and Alahi, Alexandre and Fei-Fei, Li},
  booktitle=ECCV,
  pages={694--711},
  year={2016},
  organization={Springer}
}

@inproceedings{lpips,
  title={The unreasonable effectiveness of deep features as a perceptual metric},
  author={Zhang, Richard and Isola, Phillip and Efros, Alexei A and Shechtman, Eli and Wang, Oliver},
  booktitle=CVPR,
  pages={586--595},
  year={2018}
}

@article{sun2024implicit,
  title={Recent advances in implicit representation-based 3d shape generation},
  author={Sun, Jia-Mu and Wu, Tong and Gao, Lin},
  journal={Visual Intelligence},
  volume={2},
  number={1},
  pages={9},
  year={2024},
  publisher={Springer}
}

@article{wu2024recent,
  title={Recent advances in 3d gaussian splatting},
  author={Wu, Tong and Yuan, Yu-Jie and Zhang, Ling-Xiao and Yang, Jie and Cao, Yan-Pei and Yan, Ling-Qi and Gao, Lin},
  journal={Computational Visual Media},
  volume={10},
  number={4},
  pages={613--642},
  year={2024},
  publisher={TUP}
}

@article{chen2024deepfacereshaping,
  title={DeepFaceReshaping: Interactive deep face reshaping via landmark manipulation},
  author={Chen, Shu-Yu and Jiang, Yue-Ren and Fu, Hongbo and Han, Xinyang and Liu, Zitao and Li, Rong and Gao, Lin},
  journal={Computational Visual Media},
  volume={10},
  number={5},
  pages={949--963},
  year={2024},
  publisher={TUP}
}

@article{gao2025learning,
  title={Learning multi-grained interpretable latent representation for {3D} face manipulation},
  author={Gao, Wenjing and Ji, Naye and Li, Xi and Yu, Dingguo},
  journal={Computational Visual Media},
  year={2025},
  publisher={TUP}
}

@article{zhang2025magictalk,
  title={MagicTalk: Implicit and explicit correlation learning for diffusion-based emotional talking face generation},
  author={Zhang, Chenxu and Wang, Chao and Zhang, Jianfeng and Xu, Hongyi and Song, Guoxian and Xie, You and Luo, Linjie and Tian, Yapeng and Feng, Jiashi and Guo, Xiaohu},
  journal={Computational Visual Media},
  year={2025},
  publisher={TUP}
}

@inproceedings{3dgsblendshapes,
  title={3d gaussian blendshapes for head avatar animation},
  author={Ma, Shengjie and Weng, Yanlin and Shao, Tianjia and Zhou, Kun},
  booktitle=SIGGRAPH,
  pages={1--10},
  year={2024}
}

@inproceedings{rgbavatar,
  title={Rgbavatar: Reduced gaussian blendshapes for online modeling of head avatars},
  author={Li, Linzhou and Li, Yumeng and Weng, Yanlin and Zheng, Youyi and Zhou, Kun},
  booktitle=CVPR,
  pages={10747--10757},
  year={2025}
}

@inproceedings{realtimehighfidelity,
  title={Real-time high-fidelity Gaussian human avatars with position-based interpolation of spatially distributed MLPs},
  author={Zhan, Youyi and Shao, Tianjia and Yang, Yin and Zhou, Kun},
  booktitle=CVPR,
  pages={26297--26307},
  year={2025}
}

@article{nerffacelighting,
  title={Nerffacelighting: Implicit and disentangled face lighting representation leveraging generative prior in neural radiance fields},
  author={Jiang, Kaiwen and Chen, Shu-Yu and Fu, Hongbo and Gao, Lin},
  journal=TOG,
  volume={42},
  number={3},
  pages={1--18},
  year={2023},
  publisher={ACM New York, NY, USA}
}

@article{hmdgaze,
  title={{3D} face reconstruction and gaze tracking in the HMD for virtual interaction},
  author={Chen, Shu-Yu and Lai, Yu-Kun and Xia, Shihong and Rosin, Paul L and Gao, Lin},
  journal=TMM,
  volume={25},
  pages={3166--3179},
  year={2022},
  publisher={IEEE}
}
}


\end{document}